\documentclass[useAMS,usenatbib,usegraphicx]{mn2e}

\usepackage{graphicx}    
\usepackage{multirow} 
\usepackage{tabularx} 
\usepackage[english]{babel}
\usepackage[utf8]{inputenc}
\usepackage{times}

\newcommand{\mnras}{MNRAS}

\newcommand{\aap}{A\&A}

\title[A synchrotron jet from a post-AGB star.]{A synchrotron jet from a post-AGB star.}
\author[A. F. Pérez-Sánchez et al.]{A. F. Pérez-Sánchez$^{1}$\thanks{E-mail:aperez@astro.uni-bonn.de}, W. H. T. Vlemmings$^{2}$, D. Tafoya$^{2}$ and J. M. Chapman$^{3}$\\
$^{1}$Argelander Institute for Astronomy, University of Bonn, Auf dem Hügel 71, 53121 Bonn, Germany\\
$^{2}$Onsala space observatory, Chalmers university of technology, 439 92, Onsala, Sweden.\\
$^{3}$CSIRO Astronomy and Space Science, Australia Telescope National Facility, PO Box 76, Epping, NSW 1710, Australia}

\date{Release 2013 XXxxxx XX}

\pagerange{\pageref{firstpage}--\pageref{lastpage}} \pubyear{2013}

\begin{document}
\label{firstpage}

\maketitle

\begin{abstract}
 The evolution of low- and intermediate-initial-mass
  stars beyond the asymptotic giant branch (AGB)
 remains poorly
  understood. High-velocity outflows launched shortly after the AGB
  phase are thought to be the primary 
 shaping mechanism of bipolar
  and multipolar planetary nebulae (PNe). However, little is known
  about the launching and driving mechanism for these jets, whose
  momentum and energy often far exceed the energy that can be
  provided by radiation pressure alone.  Here, we report direct
  evidence 
 of a magnetically collimated jet shaping the bipolar
  morphology of the circumstellar envelope of a post-AGB star. We
  present radio continuum observations of the post-AGB star IRAS
  15445$-$5449 (OH 326.5$-$0.4) which has water masers tracing a
  fast bipolar outflow. Our observations confirm the earlier
  observed steep negative spectral index of the spectral energy
  distribution (SED) above $\sim3$~GHz, and resolve, for the first
  time, the emission to originate from a radio jet, proving the
  existence of such jets around a post-AGB star. The SED is
  consistent with a synchrotron jet embedded in a sheath of thermal
  electrons. We find a close correspondence between the extent
  and direction of the synchrotron jet and the bipolar shape of the
  object observed at other wavelenghts, suggesting that the jet is
  responsible for the source morphology. The jet is collimated
  by a magnetic field of the order of mG at almost 7000 AU from the
  central star. We recover observations from the ATCA archive that
  indicate that the emission measure of the thermal component has
  increased by a factor of three between 1998 and 2005 after which it
  has remained constant. The short timescale evolution of the radio
  emission suggests a short lifetime for the jet. The observations of
  a synchrotron jet from a post-AGB star with characteristics similar
  to those from protostars and young stellar objects, for
  instance, suggest that magnetic launching and collimation is a
  common feature of astrophysical jets.
\end{abstract}


\section{Introduction}
Synchrotron radiation is commonly detected towards relativistic jets emerging in high-energy astrophysical sources
such as active galactic nuclei and quasars. It also has been detected towards a magnetized jet from a young stellar object (YSO) \citep{charly}, where owing
to density and temperature conditions, relativistic outflows were thought to be unlikely. The mechanism that leads to synchrotron emission 
involves the interaction between relativistic particles moving in a region where the dynamics are controlled by a magnetic field. Highly 
collimated outflows, similar to those seen toward YSOs, have also been found in the vicinity of post-Asymptotic Giant Branch (post-AGB) stars \citep{buja}. If these high-velocity outflows 
are launched under the influence of strong magnetic fields, synchrotron emission could be expected. However, to date, no direct evidence of synchrotron radio emission exists from these 
high-velocity sources. 

Post-AGB stars are thought to represent the group of stars that recently left the asymptotic giant branch (AGB) phase and will develop
into a planetary nebula (PN) \citep{VanW}. During the evolution of low- and intermediate-initial-mass stars ($\sim$ 1-8 M$_{\odot}$) on the AGB, high mass-loss rates
($10^{-7}<\dot{M}<10^{-4}$~M$_{\odot}$~yr$^{-1}$) together with an acceleration mechanism, which drives the ejected material outwards, lead to the formation
of a dense and extended circumstellar envelope (CSE). The CSE expands outwards in radial directions
with constant velocity of order 10 km~s$^{-1}$. Generally, it is assumed that the mass-loss process is spherically symmetric during the AGB \citep{agbook} and that asymmetries form
in a short timescale before entering the PNe phase (e.g.~\citealt{Sahai}). Therefore, the post-AGB is a key phase for the understanding of the evolution of the CSE, where the mechanism(s) 
responsible for shaping asymmetric PNe must become important. 
Binary systems, large-scale magnetic fields and interactions with substellar companions are among the proposed mechanisms that can generate aspherical CSEs \citep{balick}.
The evolution of the CSEs beyond the AGB often involves the interaction between a fast collimated wind, which could be created during the very last thermal pulses of the central star, and 
the steadily expanding CSE formed during the AGB. Strong evidence of this interaction and of the generation of strong shocks within the CSE, is the 
detection of post-AGB sources with water maser emission spread over unusually large velocity ranges ($\geq 100$~km~s$^{-1}$) the so called water fountains \citep{likkel,gomez}. 
The peculiar H$_{2}$O maser emission is thought to trace regions that have been swept up by a high-velocity outflow. Polarized H$_{2}$O maser emission has also been 
detected towards water fountains, suggesting that the high-velocity outflow of at least one of the water fountain sources is collimated by the magnetic field \citep{VlemmingsNat}. Nevertheless,
the lauching mechanism of high-velocity outflows in post-AGB stars, as well as the origin of the collimating magnetic field, are still under debate.  

One of the 20 detected water fountains is IRAS 15445$-$5449, which has
been classified as an evolved post-AGB star according to the MSX
two-color diagram \citep{sevemsx,deacone}. Observations have
  revealed H$_{2}$O maser emission spread over $\approx
  100$~km~s$^{-1}$, with the spectral features redshifted with respect
  to the systemic velocity of the source.  Based on both satellite-
  and main-line OH maser observations, \citet{deacontwo} suggested
  that $v_{lsr}\approx -150$~km~s$^{-1}$. We note however, that
  the source has also been classified as a massive YSO candidate in
  the RMS survey, where radio continuum was detected at 3~cm
  \citep{Urquhart}, a classification that could be supported by the
  detection of relatively strong line emission of $^{13}$CO$(1-0)$
  centered at $-44$~kms$^{-1}$ \citep{UrquhartCO}. However, if the
  $^{13}$CO is associated with IRAS 15445$-$5449, the central velocity
  of the line emission will define the $v_{lsr}$ of the source, and in
  that case, both the H$_{2}$O and OH maser emission would arise from
  a blue-shifted outflow. Specifically, this would imply that the OH
  emission centered at $v_{lsr}\approx -150$~km~s$^{-1}$ and spread
  over $\sim60$~km~s$^{-1}$, originates in an outflow with
  velocity $>200$~km~s$^{-1}$. This would correspond to the
  highest velocity Galactic OH maser and is very
  unlikely. Additionally, strong Galactic CO emission is known to
  occur around $-40$~km$^{-1}$, leading us to conclude that the
  $^{13}$CO emission is likely Galactic in origin and not associated
  with 15445$-$5449.  Although a YSO classification cannot yet be
  ruled out, its classification as a post-AGB star is more likely.

The projected spatial distribution of the detected H$_{2}$O maser
features resemble a bow-shock structure at the red-shifted lobe, which
is likely caused by a collimated outflow that pierces the
steady-expanding CSE of the AGB phase \citep{mine}. Additionally,
mid-infrared images confirm the bipolar morphology of this source
\citep{lagadec}. Finally, strong and seemingly non-thermal radio
emission was detected and thought to arise from either the star
itself or a shocked interaction region between the AGB envelope and
a fast wind \citep{bains}. Here we
report observations indicating the radio continuum to originate from
a jet.

\section{Observations and results}

\subsection{Recent and archive observations}

We performed observations of IRAS~15445$-$5449 with the Australia Telescope Compact Array (ATCA) on 2012 September 2. 
The 12-h observation run was carried out with the 6A array configuration, using 2~GHz bandwidths at 1.3~cm, 
3~cm, 6~cm, and 16~cm. The observations with the 1.3~cm and 16~cm bands were carried out using the Compact Array Broadband Backend (CABB) 
mode 1~M-0.5~k, whereas the 3~cm and 6~cm bands were set up simultaneously with CABB mode 64~M-32~k. The four bands were set up in
full polarization mode. The calibration and the imaging of the data were done using the package MIRIAD \citep{miriad}. 
Bandpass and flux calibration were performed on the standard calibrator 1934$-$638 
and phase calibration was performed on 1613$-$586. 
The flux densities for 1934$-$638 were taken from the available models 
of this calibrator. The flux densities of 1613$-$586 were determined in order to check the absolute flux calibration 
accuracy. The measured flux densities for this source are in agreement with the values presented in the ATCA calibrators database, within an 
uncertainty of less than $10\%$. The bands at 3~cm, 6~cm, and 16~cm were split in two sub-bands each, in order
to confirm a potential steep spectral index over the large fractional bandwidth at these wavelength. 
After calibration, the imaging of the source was performed using multifrequency synthesis. The different maps were deconvolved
using the MIRIAD task MFCLEAN, with which Stokes I, Q, U and V images were produced. No linear or circular polarization was detected. Maps of smaller frequency ranges were created in order to test 
if the lack of linear polarization was due to a large rotation measure across the individual bands, but the emission was found to be unpolarized. 
We also retrieved radio continuum observations of IRAS
15445$-$5449 from the Australia Telescope Online Archive, the results
of which are included in Table~\ref{tb:two}.

\subsection{Maps and spectral energy distribution}

\begin{figure}
\begin{minipage}{82mm}
  \includegraphics[width=70mm]{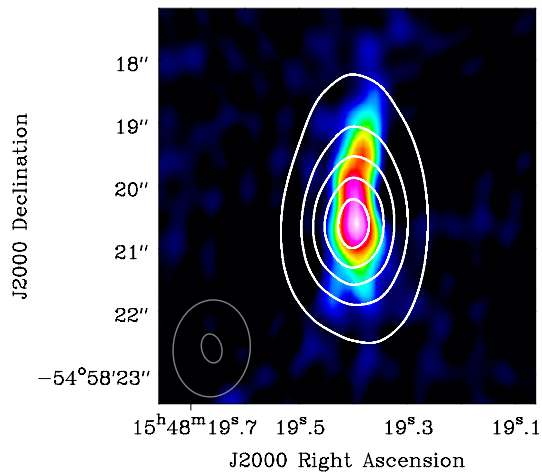}
  \caption{Radio continuum map of IRAS 15445$-$5449 at 22.0~GHz (color) and 5~GHz (contours). 
           The peak-flux density of the 22.0~GHz image is 
           $3.65$~mJy/beam, while $\sigma=5.0\times10^{-2}$~mJy. The contour lines at 5~GHz are drawn at 
           $10, 30, 50, 70,$~and~$90\%$ of the peak flux of $16.6$~mJy/beam. 
           The beam size for the 22.0 and 5~GHz observations are drawn in the bottom left corner.}
\label{fig:map}
\end{minipage}
\end{figure}

\begin{figure}
  \includegraphics[width=82mm]{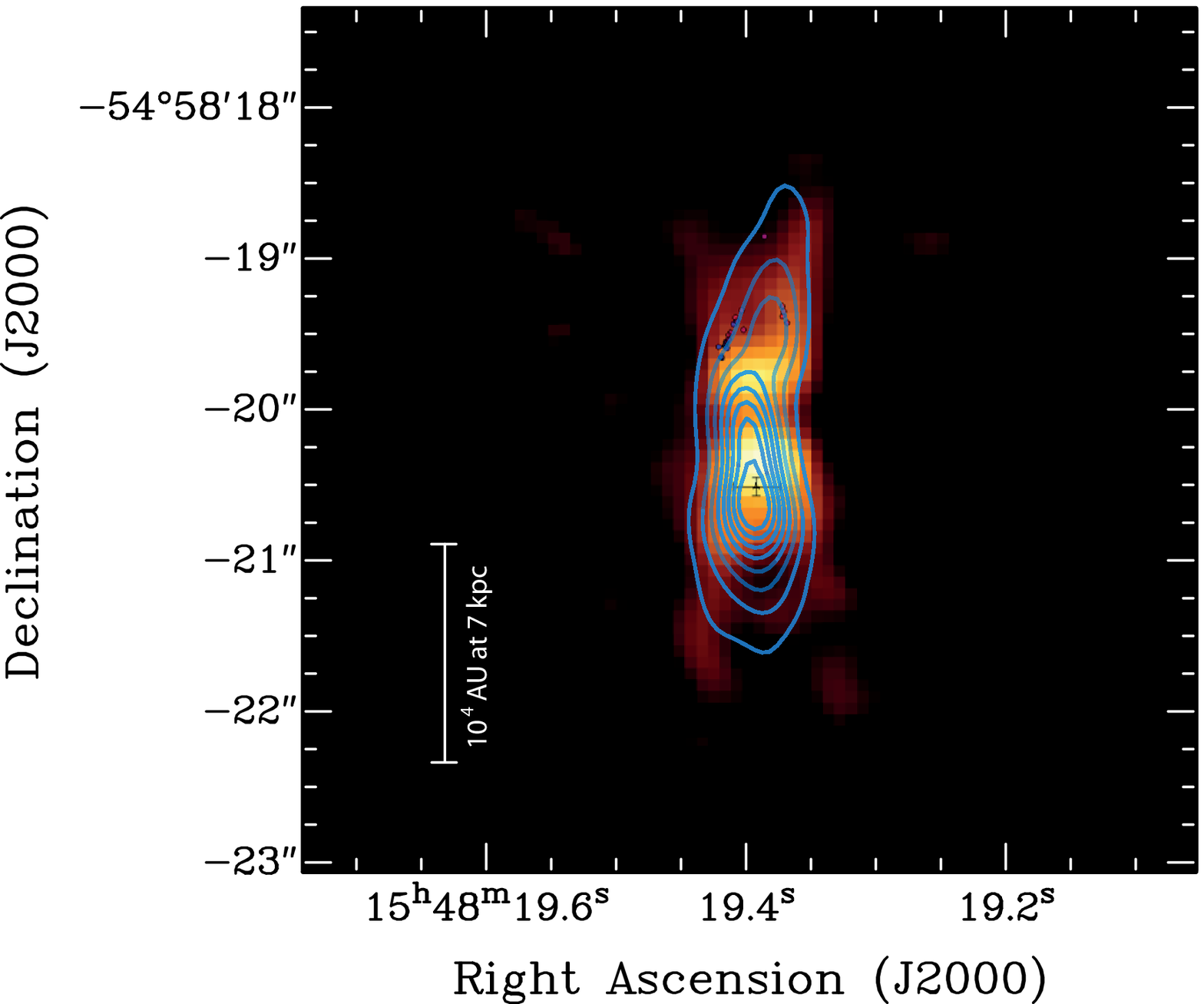}
  \caption{2012 radio continuum map of IRAS 15445$-$5449 at 22.0 GHz (contours) overlaid on the mid-infrared VLTI image \citep{lagadec}, the H$_2$O
           masers (colored symbols) observed in the redshifted lobe of the high-velocity outflow \citep{mine} and the radio continuum position
          (solid triangle with error bars) determined in the 2005 epoch \citep{bains}. The contours are drawn from $10\sigma$ at intervals of
          $10\sigma$. The mid-infrared image has been shifted to match the observed outflow for illustration, as the positional uncertainty of
          the mid-infrared observations, with an original centre of RA~$15^{h}48^{m}19^{s}.42$ and Dec~$-54^{\circ}58'20".10$, is 2 arcseconds. }
\label{fig:compo}
\end{figure}

The derived flux densities, the beam size, and the rms of each map of IRAS 15445$-$5449 are listed in Table~\ref{tb:two} for our observations in 2012, the observations
in 2005 reported by \citet{bains}, and the data taken from the ATCA archive. The source was detected in 1998 at 0.84~GHz (D. Hunstead, private comunication; \citet{Molonglo}). It was also
detected during another survey between 2002 and 2004 at 8.6~GHz, with a flux density of 11.9~mJy \citep{Urquhart}. The observed spectral energy distribution (SED) of IRAS 15445$-$5449 at 
radio frequencies is characterized by a turn-over around $\nu_{c}\approx 3$~GHz (Fig.~\ref{fig:spectra}) with a steep negative spectral index at higher frequencies. 
Between $5.0$~GHz and $22$~GHz the continuum emission has a spectral index $\alpha=-0.56$.
At 22~GHz we resolved the dense continuum emission along the north-south direction (Fig.~\ref{fig:map}) extending about 1.9 arcsec, which is consistent with the bipolar 
morphology seen in the mid-infrared image of the source (Fig.~\ref{fig:compo}). 

\section{Analysis}

\begin{table*}
 \centering
\begin{minipage}{160mm}
   \caption{Flux density of IRAS 15445$-$5449.}
   \begin{tabular}{ c  c  c  |  c  c  c | c  c  c}
\hline
 \multicolumn{3}{c}{Epoch 2012$^{a}$} & \multicolumn{3}{c}{Epoch 2005$^{b}$} & \multicolumn{3}{c}{Epoch 1998/99}\\
\hline
 Frequency &        Flux        &  Beam            &  Frequency  &        Flux        &   Beam            & Frequency &     Flux     & Beam \\
  (GHz)    &        (mJy)       & (arcsec$^{2}$)    &   (GHz)     &        (mJy)       & (arcsec$^{2}$)    &   (GHz)    &    (mJy)     & (arcsec$^{2}$)\\
\hline
    -     &        -           &      -            &     -       &        -           &       -           &    0.84   & 32.5$\pm$4.1$^{c}$  &   -\\
   2.1    &  20.36$\pm$0.14    & 5.6$\times$3.0    &     -       &        -           &       -           &    1.4    & 23.0$\pm$5.3 &  7.6$\times$4.9$^{d}$\\
   2.5    &  21.72$\pm$0.05    & 4.7$\times$2.6    &     -       &        -           &       -           &    2.5    & 31.2$\pm$6.2 &  3.3$\times$3.0$^{e}$\\
   5.0    &  25.41$\pm$0.04    & 1.9$\times$1.5    &    4.8      &  22.81$\pm$1.11    & 5.6$\times$2.1    &    4.8    & 18.4$\pm$2.4 &  2.0$\times$1.6$^{e}$\\
   5.8    &  23.59$\pm$0.05    & 1.6$\times$1.3    &     -       &        -           &       -           &     -     &      -       & - \\
   8.5    &  18.45$\pm$0.05    & 1.1$\times$0.9    &    8.6      &  18.67$\pm$0.47    & 3.0$\times$1.2    &    8.6    & 10.7$\pm$2.3 &  1.2$\times$0.8$^{e}$\\
   9.3    &  17.70$\pm$0.03    & 1.0$\times$0.8    &     -       &        -           &       -           &     -     &     -         & - \\
   22.0   &  12.31$\pm$0.03    & 0.48$\times$0.34  &     -       &        -           &       -           &     -     &     -         & - \\
\hline
\label{tb:two}
\end{tabular}
\\$^{a}$ This paper, $^{b}$ \citet{bains}, $^{c}$ Observations carried out on 1998 June 13 (D. Hunstead, private comunication; \citet{Molonglo}), $^{d}$ Observations taken on 1999 September 2 (ATCA archive), $^{e}$ Observations carried out on 1998 November 9$-$14 \citep{deacontwo}.
\end{minipage}
\end{table*}

\subsection{Model}
The shape of the SED indicates that the radio continuum is a superposition of a foreground thermal component, which is optically thick for $\nu < \nu_{c}$, 
and a non-thermal component, which dominates the radio continuum flux when the thermal component becomes optically thin, i.e., for $\nu \geq \nu_{c}$.
We model the observations as arising from non-thermal emission in a cylindrical region surrounded by a sheath of thermal electrons with constant 
density ($n_{e}=3.5\times 10^{4}$~cm$^{-3}$) and temperature ($T_{e}=6000$~K) and a thickness of approximately $1000$~AU. The thickness is limited by the unresolved width of the emission 
at 22~GHz. The brightness temperature of the non-thermal jet implies its width to be less than $500$~AU. For this configuration the emission has two regimes: at low frequencies 
the thermal emission becomes optically thick, which causes the emission from the non-thermal electrons to be absorbed, and the SED to exhibit a positive spectral index. 
On the other hand, since at high frequencies the thermal emission becomes optically thin, the contribution from the non-thermal electrons becomes dominant, resulting in the negative spectral index.
Although the spectral index indicates synchroton emission, polarized emission was not detected. This is a direct consequence of depolarization due to the thermal electrons surrounding 
the region where the synchrotron emission arises.

\subsection{Fermi acceleration and magnetic field}

Our best fit to the observed SED implies a spectral index $\alpha=-0.68\pm 0.01$ for the non-thermal component.
As the stellar temperature is approximately $12000$~K \citep{bains}, 
ionization by the star is unlikely to provide many free electrons. But, if we assume a strong shock (J-shock) 
between the magnetically collimated outflow and the slow AGB wind, then ionized material will propagate downstream 
throughout the shock-front. Eventually, a fraction of the electrons will be accelerated to relativistic 
speeds through the Fermi shock acceleration mechanism. Hence, Fermi accelerated electrons interacting with 
the magnetic field lines that collimate the outflow generate the observed 
non-thermal emission.  Within this framework, we can assume that the total internal energy of the shocked region is split between the particles (electrons and protrons) and the magnetic field.  
When the total energy density is minimized with respect to the magnetic field, the energy density of the particles becomes nearly equal to the energy density of the magnetic field, and 
we can assume equipartition of the energy. This assumption enables us to estimate the strength of the minimum-energy magnetic field interacting with the relativistic electrons. 
This minimum-energy magnetic field and the minimum total energy can be calculated from 

\begin{eqnarray}
B_{min} & = & [4.5c_{12}(1+k)L/\phi]^{2/7}R^{-6/7}\ \rmn{G,\ \ and}\\
E_{min} & = & c_{13}[(1+k)L]^{4/7}\phi^{3/7}R^{9/7} \rmn{erg}, 
\end{eqnarray}

where $R$ is the size of the source, $L$ is the integrated radio
luminosity, $k$ is the ratio between the energy of heavy particles
(protons) and the electrons, $\phi$ is the volume filling 
factor of the emitting region, and $c_{12}$ and $c_{13}$ are functions of both the spectral index, and the maximum and minimum frequencies considered for the integration of the spectral 
energy distribution \citep{RAbook}. The distance to the source is not known, and only its (near) kinematic distance has been reported $D= 7.1$~kpc. From the measured extent we 
estimate that the size of the source is $R=6.74\times 10^{3}~[D/7.1~$kpc$]$~AU. Integrating the radio luminosity between $\nu_{min}=10^{7}$~Hz and $\nu_{max}=10^{11}$~Hz, assuming a spectral 
index of $\alpha=-0.68$, a volume filling factor $\phi=0.004[D/7.1~$kpc$]^{-2}$ (for a cylinder of 500~AU radius), and $k=40$ (which is an appropriate value for electrons undergoing 
Fermi shock acceleration in a non-relativistic jet \citep{beck}), we obtain $B_{min}=5.43~[D/7.1~$kpc$]^{-2/7}$~mG and $E_{min}=4.75\times10^{43}~[D/7.1~$kpc$]^{3/7}$~erg.
Since we have not detected any linear polarization, we cannot determine if the synchrotron radiation traces either the poloidal ($B_{r}$) or the toroidal ($B_{\phi}$) component of the 
magnetic field. However, since $B_{r} \propto r^{-2}$, this component can likely be neglected at $r > 1000$~AU compared to the toroidal field, which has $B_{\phi} \propto r^{-1}$.  Therefore, 
assuming that the measured magnetic field strength corresponds to the toroidal component at $R=6.74\times 10^{3}~[D/7.1~$kpc$]$~AU, the magnetic field at $R_{\star}=2$~AU 
is $B_{\phi}\approx 18.3~[D/7.1~$kpc$]^{1/7}$~G, which is similar to the extrapolated magnetic field strengths towards other (post-)AGB stars from maser observations \citep[e.g.][]{VlemmingsNat}.

\begin{figure}
\begin{minipage}{82mm}
  \includegraphics[width=75mm]{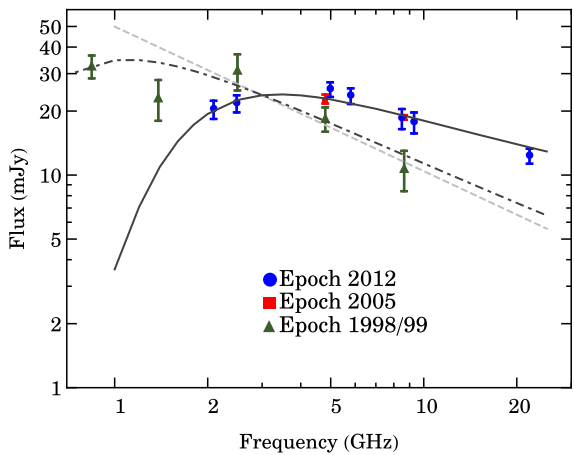}
  \caption{Spectral energy distribution of the radio continuum of IRAS 15445$-$5449 for three different epochs: 1998/99 (references listed in Table \ref{tb:two}),
           2005 reported by \citet{bains}, and our observation in 2012. The solid line (epoch 2012) and the dot-dashed line (epoch 1998/99) are the results of
           our models, that fit the observations. We consider a synchrotron jet surrounded by a sheath of thermal electrons. 
           Assuming that the synchrotron emission remained constant between 1998 and 2012, our model suggest that the 
           electron density of the region surrounding the synchrotron emitting region has increased by a factor of two between 1998 and 2005, and remained
           stable since then. The grey-dashed line indicates the synchrotron component. The best fit yielded a spectral index of $\alpha=-0.68\pm 0.01$ for the synchrotron 
           component.}
\label{fig:spectra}
\end{minipage}
\end{figure}  

\subsection{Variability and lifetime}
A negative spectral index was obtained from the archive observations towards IRAS 15445$-$5449 (Table~\ref{tb:two}).  The spectral index below $13$~cm, of the SED observed in 1998/1999, 
is $\alpha=-0.85\pm 0.05$, with a turn-over frequency shifted toward a
lower frequency compared with the SED observed in 2012. 
Additionally, radio continuum observation at $3$~cm and $6$~cm carried out in 2005 yielded a spectral index $\alpha=-0.34\pm 0.24$ \citep{bains}. Because it is a single data point taken 
between epochs 1998/99 and 2005, the observation at $3$~cm reported by \citet{Urquhart}, which yielded a similar flux density to that reported by \citet{bains}, is not 
included in our fit. Since the time between the two observations is negligible 
compared with the synchrotron life time of relativistic electrons, we can assume that the flux of the synchrotron 
emission remained nearly constant between 1998 and 2012 and fit all observations considering a single synchrotron component and an increasing emission
measure of the surrounding sheath of thermal electrons (Fig.~\ref{fig:spectra}). An increase of the emission measure by a factor of 3 between 1998 and 2005, and stable thereafter, 
can fit all three observational epochs. This implies that, assuming the dimensions of the surrounding sheath of thermal electrons remained the same, the electron density increased by a 
factor of two in seven years, likely by the same shock-ionization process that produces the electrons that are accelerated to relativistic velocities by the Fermi mechanism. In recent years, 
the production rate of thermal electrons is in equilibrium with their recombination rate. The stability of the non-thermal component over almost 15 years indicates that this 
emission is unlikely to originate from the colliding winds of binary systems such as observed around, for example, binary Wolf-Rayet stars \citep[e.g.][]{chapman}. 

The rapid initial increase of emission measure between 1998 and 2005 likely implies the jet, responsible for the ionizing fast shocks, to have been launched only shortly before. 
Additionally, as the synchrotron flux depends on the effectiveness of the Fermi shock acceleration mechanism, once the shock front reaches outer layers where the strength of 
$B_{\phi}$ and the density of the CSE decrease, the synchrotron component would no longer be observable. This also suggest that the magnetically collimated outflow was launched recently. 
Thus, the lifetime of the synchrotron radiation toward post-AGB sources will most likely be determined by the time of propagation of the collimated outflow throughout regions of the 
CSE where the shock conditions enable the Fermi shock acceleration mechanism. Consequently, the time scale for the synchrotron radiation would at most be a few hundred years. 
Depending on the initial mass of the star, the synchrotron radiation time scale would be shorter than, or almost comparable with, the time scale of the post-AGB phase 
\citep[][and references therein]{VanW}. The fact that the source is a water fountain suggests that, in principle, this kind of source would have conditions required to trigger the 
Fermi shock acceleration mechanism. Nevertheless, a larger sample of confirmed water fountains and further observations are necessary to directly correlate both phenomena.

\subsection{Implications for post-AGB outflows}
The physical processes responsible for shaping the asymmetrical envelopes observed towards PNe have been subject of intense debate along 
the last three decades. Recently, it has been suggested that the bipolar structures observed towards post-AGB stars and young PNe are 
associated with low-density axisymmetric regions that are illuminated by a central star obscured by a dense equatorial torus \citep[e.g.][]{koning}. 
These low-density regions are assumed to be formed as a result of the propagation of high-momentum, collimated outflows that emerge from the inner regions of 
the CSE, creating cavities along the axis defined by its propagation direction. Nevertheless, the actual formation process of such cavities is not yet clear. 
On the other hand, observations of molecular outflows traced by CO lines towards a large sample of post-AGB stars revealed the existence of very fast 
collimated outflows \citep{buja}. The momentum carried by most of these fast, highly collimated outflows cannot be explained considering a 
radiatively driven wind only, but the minimum energy calculated for the jet of IRAS 15445$-$5449 is within the typical range of the kinetic energy measured from the 
observation of these molecular outflows ($10^{42}-10^{46}$~erg). 
Magnetohydrodynamical simulations have shown that magnetic fields can be an important agent in the collimation of the outflows 
observed towards PNe \citep{garcia}. Dust polarization observations have been carried out in order to trace the magnetic field morphology.
But so far the observational study of the properties of magnetic
fields towards AGB and post-AGB stars and PNe has relied on the detection of dust polarization and polarized 
maser emission arising in their CSEs and high-velocity outflows (e.g. \citet{sabin,VlemmingsNat}). 
Our results provide strong observational evidence that indicates that the magnetic field is an important source of energy and is thus of great 
importance for the launching and driving mechanisms of the high-velocity jets from post-AGB stars. Furthermore, it will now be possible to directly test which class of 
magnetic launching models fits the observations \citep[e.g][]{huarte}. 

\section{Discussion}

For the first time, we resolve a synchrotron jet towards a post-AGB star. The resolved radio continuum emission is
consistent with the bipolar morphology of IRAS 15445$-$5449 observed at the infrared, reported by \citet{lagadec}, and suggests the jet is responsible for shaping the CSE.
Although both theoretical models and previous H$_{2}$O maser observations have been used to infer the presence of magnetically
collimated outflows towards post-AGB stars, our result represents a direct observational evidence and the first unambiguous
proof that magnetic fields are a key agent to explain the asymmetries observed towards PNe.   
Still, the source of the stellar magnetic field remains unclear. A large scale magnetic field could arise from (convective) dynamo action in a single star
or require a binary (or planetary) system to be maintained. In one of the binary scenarios, the bipolar
outflows are launched from a low-mass companion accreting mass ejected by the more evolved star.
In this case, the collimation of the outflow might occur via a mechanism similar to that collimating the bipolar outflows from protostars, for example a disc-wind or an X-wind (\citealt{Blandford,shu}). 
Our results cannot yet discern which is the most likely scenario for IRAS 15445$-$5449, although a hint of curvature might point to a binary ejection mechanism. 
Finally, our detection of synchotron radiation towards IRAS 15445$-$5449 also demonstrates that the conditions for the Fermi
shock acceleration of electrons can be attained at the final stages of the evolution of intermediate-initial-mass stars.

\section{Acknowledgments}
This research was supported by the Deutsche Forschungsgemeinschaft
(DFG) through the Emmy Noether Research grant VL 61/3-1. The authors
also thank the referee Albert Zijlstra as well as Bruce Balick for
insightful comments that improved the manuscript. The Australia
Telescope Compact Array is part of the Australia Telescope National
Facility which is funded by the Commonwealth of Australia for
operation as a National Facility managed by CSIRO. This paper includes
archived data obtained through the Australia Telescope Online Archive
(http://atoa.atnf.csiro.au).

\label{lastpage}

\begin{thebibliography}{99}
  \bibitem[\protect\citeauthoryear{Bains et al.}{2009}]{bains} Bains, I., Cohen, M., Chapman, J.~M., Deacon, R.~M., \& Redman, M.~P.\ 2009, MNRAS, 397, 1386

  \bibitem[\protect\citeauthoryear{Balick et al.}{2002}]{balick} Balick, B., \& Frank, A.\ 2002, Ann. Rev. A\&A, 40, 439 

  \bibitem[\protect\citeauthoryear{Beck \& Krause}{2005}]{beck} Beck, R., \& Krause, M.\ 2005, Astronomische Nachrichten, 326, 414

  \bibitem[\protect\citeauthoryear{Blandford \& Payne}{1982}]{Blandford} Blandford, R.~D., \& Payne, D.~G.\ 1982, MNRAS, 199, 883

  \bibitem[\protect\citeauthoryear{Bujarrabal et al.}{2001}]{buja} Bujarrabal, V., Castro-Carrizo, A., Alcolea, J., \& S{\'a}nchez Contreras, C.\ 2001, A\&A, 377, 868 

  \bibitem[\protect\citeauthoryear{Carrasco-González et al.}{2010}]{charly} Carrasco-Gonz{\'a}lez, C., Rodr{\'{\i}}guez, L.~F., Anglada, G., et al.\ 2010, Science, 330, 1209

  \bibitem[\protect\citeauthoryear{Chapman et al.}{1999}]{chapman} Chapman, J.~M., Leitherer, C., Koribalski, B., Bouter, R., \& Storey, M.\ 1999, ApJ, 518, 890

  \bibitem[\protect\citeauthoryear{Deacon et al.}{2004}]{deacone} Deacon, R.~M., Chapman, J.~M., \& Green, A.~J.\ 2004, ApJS, 155, 595

  \bibitem[\protect\citeauthoryear{Deacon et al.}{2007}]{deacontwo} Deacon, R.~M., Chapman, J.~M., Green, A.~J., \& Sevenster, M.~N.\ 2007, ApJ, 658, 1096    

  \bibitem[\protect\citeauthoryear{Garc\'ia-Segura et al.}{1999}]{garcia} Garc{\'{\i}}a-Segura, G., Langer, N., R{\'o}{\.z}yczka, M., \& Franco, J.\ 1999, ApJ, 517, 767  

  \bibitem[\protect\citeauthoryear{Gómez et al.}{2011}]{gomez} G{\'o}mez, J.~F., Rizzo, J.~R., Su{\'a}rez, O., et al.\ 2011, ApJL, 739, L14

  \bibitem[\protect\citeauthoryear{Habing \& Olofsson}{2003}]{agbook} Habing, H.~J., \& Olofsson, H.\ 2003, Asymptotic giant branch stars,  Astronomy 
    and astrophysics library, New York, Berlin: Springer, 2003.

  \bibitem[\protect\citeauthoryear{Huarte-Espinosa et al.}{2012}]{huarte} Huarte-Espinosa, M., Frank, A., Blackman, E.~G., et al.\ 2012, ApJ, 757, 66 

  \bibitem[\protect\citeauthoryear{Koning, Kwok \& Steffen}{2013}]{koning} Koning, N., Kwok, S., \& Steffen, W.\ 2013, ApJ, 765, 92 

  \bibitem[\protect\citeauthoryear{Lagadec et al.}{2011}]{lagadec} Lagadec, E., Verhoelst, T., M{\'e}karnia, D., et al.\ 2011, MNRAS, 417, 32

  \bibitem[\protect\citeauthoryear{Likkel \& Morris}{1988}]{likkel} Likkel, L., \& Morris, M.\ 1988, ApJ, 329, 914

  \bibitem[\protect\citeauthoryear{Murphy et al.}{2007}]{Molonglo} Murphy, T., Mauch, T., Green, A., et al.\ 2007, \mnras, 382, 382 

  \bibitem[\protect\citeauthoryear{Pacholczyk}{1970}]{RAbook} Pacholczyk, A.~G.\ 1970, Series of Books in Astronomy and Astrophysics, San Francisco: Freeman, 1970,

  \bibitem[\protect\citeauthoryear{Pérez-Sánchez, Vlemmings \& Chapman}{2011}]{mine} P{\'e}rez-S{\'a}nchez, A.~F., Vlemmings, W.~H.~T., \& Chapman, J.~M.\ 2011, MNRAS, 418, 1402

  \bibitem[\protect\citeauthoryear{Sabin et al.}{2007}]{sabin} Sabin, L., Zijlstra, A.~A., \& Greaves, J.~S.\ 2007, \mnras, 376, 378 

  \bibitem[\protect\citeauthoryear{Sahai et al.}{2007}]{Sahai} Sahai, R., Morris, M.,  S{\'a}nchez Contreras, C., \& Claussen, M.\ 2007, AJ, 134, 2200

  \bibitem[\protect\citeauthoryear{Sault, Teuben \& Wright}{1995}]{miriad} Sault, R.~J., Teuben, P.~J., \& Wright, M.~C.~H.\ 1995, Astronomical Data Analysis Software and Systems IV, 77, 433 

  \bibitem[\protect\citeauthoryear{Sevenster}{2002}]{sevemsx} Sevenster, M.~N.\ 2002, AJ, 123, 2772 

  \bibitem[\protect\citeauthoryear{Shu et al.}{1994}]{shu} Shu, F., Najita, J.,Ostriker, E., et al.\ 1994, ApJ, 429, 781

  \bibitem[\protect\citeauthoryear{Urquhart et al.}{2007}]{Urquhart} Urquhart, J.~S., Busfield, A.~L., Hoare, M.~G., et al.\ 2007, \aap, 461, 11 

  \bibitem[\protect\citeauthoryear{Urquhart et al.}{2007}]{UrquhartCO} Urquhart, J.~S., Busfield, A.~L., Hoare, M.~G., et al.\ 2007, \aap, 474, 891 

  \bibitem[\protect\citeauthoryear{van Winckel}{2003}]{VanW} van Winckel, H.\ 2003, Ann. Rev. A\&A, 41, 391

  \bibitem[\protect\citeauthoryear{Vlemmings et al.}{2006}]{VlemmingsNat} Vlemmings, W.~H.~T., Diamond, P.~J., \& Imai, H.\ 2006, Nature, 440, 58 
\end{thebibliography}
\end{document}